\documentclass[runningheads]{llncs}
\usepackage{graphicx}
\begin{document}
%%%%%%%%%%%%%%%%%%%%%%%%%%%%%%%%%%%%%%%%%%%%%%%%%%%%%%%%%%%%%%%%%%%%%%%%%%%%%
%%%%%%%%%%%%%%%%%%%%%%%%%%%%%%%%%%%%%%%%%%%%%%%%%%%%%%%%%%%%%%%%%%%%%%%%%%%%%
%%%%%%%%%%%%%%%%%%%%%%%%%%%%%%%%%%%%%%%%%%%%%%%%%%%%%%%%%%%%%%%%%%%%%%%%%%%%%
%
\title{Distributed data storage for modern astroparticle physics experiments\thanks{Supported by RSF, grant no.18-41-06003}}
\author{
Alexander Kryukov\inst{1,3}\orcidID{0000-0002-1624-6131} 
\and Minh-Duc Nguyen\inst{1}\orcidID{0000−0002−5003−3623}
\and Igor Bychkov\inst{2}
\and Andrey Mikhailov\inst{2}\orcidID{0000-0001-5572-5349}
\and Alexey Shigarov\inst{2}
\and Julia Dubenskaya\inst{1}\orcidID{0000-0002-2437-4600}
}
\authorrunning{A. Kryukov et al.}
\institute{
Skobeltsyn Institute of Nuclear Physics, Lomonosov Moscow State University, Moscow 119992, Russia
\and Matrosov Institute for System Dynamics and Control Theory \\ \email{kryukov@theory.sinp.msu.ru, nguyendmitri@gmail.com, jdubenskaya@gmail.com}
     Siberian Branch of Russian Academy of Sciences, Lermontov st. 134, Irkutsk, Russia \\ \email{shigarov@icc.ru}
}
\maketitle              % typeset the header of the contribution
%%%%%%%%%%%%%%%%%%%%%%%%%%%%%%%%%%%%%%%%%%%%%%%%%%%%%%%%%%%%%%%%%%%%%%%%%%%%%
%
%%%%%%%%%%%%%%%%%%%%%%%%%%%%%%%%%%%%%%%%%%%%%%%%%%%%%%%%%%%%%%%%%%%%%%%%%%%%%
\begin{abstract}
%%%%%%%%%%%%%%%%%%%%%%%%%%%%%%%%%%%%%%%%%%%%%%%%%%%%%%%%%%%%%%%%%%%%%%%%%%%%%

The German-Russian Astroparticle Data Life Cycle Initiative is an international project launched in 2018.
The Initiative aims to develop technologies that provide a unified approach to data management, as well as to demonstrate their
applicability on the example of two large astrophysical experiments - KASCADE and TAIGA.
One of the key points of the project is the development of a distributed storage, which, on the one hand, will allow data of several experiments to be combined into a single repository with unified interface, and on the other hand, will provide data to all participants of experimental groups for multi-messenger analysis.
Our approach to storage design is based on the single write-multiple read (SWMR) model for accessing raw or centrally processed data for further analysis.
The main feature of the distributed storage is the ability to extract data either as a collection of files or as aggregated events from different sources.
In the last case the storage provides users with a special service that aggregates data from different storages into a single sample.
Thanks to this feature, multi-messenger methods used for more sophisticated data exploration can be applied.
Users can use both Web-interface and Application Programming Interface (API) for accessing the storage.
In this paper we describe the architecture of a distributed data storage for astroparticle physics and discuss the current status of our work.

\end{abstract}

%%%%%%%%%%%%%%%%%%%%%%%%%%%%%%%%%%%%%%%%%%%%%%%%%%%%%%%%%%%%%%%%%%%%%%%%%%%%%
%%%%%%%%%%%%%%%%%%%%%%%%%%%%%%%%%%%%%%%%%%%%%%%%%%%%%%%%%%%%%%%%%%%%%%%%%%%%%
\section{Introduction}
%%%%%%%%%%%%%%%%%%%%%%%%%%%%%%%%%%%%%%%%%%%%%%%%%%%%%%%%%%%%%%%%%%%%%%%%%%%%%
%%%%%%%%%%%%%%%%%%%%%%%%%%%%%%%%%%%%%%%%%%%%%%%%%%%%%%%%%%%%%%%%%%%%%%%%%%%%%

Currently, a number of experimental facilities in the field of particle astrophysics of the mega-sciences class are under construction or are already operating around the world.
Among them there are such installations as LSST~\cite{LSST,LSST2}, MAGIC~\cite{MAGIC,Ricoa16}, CTA~\cite{CTA,CTA2}, VERITAS~\cite{VERITAS}, HESS~\cite{HESS}, and others. 
These facilities collect a tremendous volume of data. 
For example the annual (reduced) raw data of the CTA project have a volume of about 4 PB.
The total volume to be managed by the CTA archive is of the order of 25 PB per year, when all data-set versions and backup replicas are considered.

In addition to the huge flow of data produced, an important feature of this class of projects is the participation of many organizations and, as a result, the distributed nature of data processing and analysis.
All this presents a real challenge to developers of the data analytics infrastructure.

To meet a similar challenge in high energy physics, the WLCG grid was deployed as part of the LHC project~\cite{WLCG}. 
This solution, on the one hand, proved to be highly efficient, but on the other hand, it turned out to be a rather heavy one requiring high administrative costs, highly qualified staff and a very homogeneous environment on which applications operate.
The success of the WLCG is based primarily on the fact that thousands of physicist users actually solve one global problem using a highly centralized system management.

Taking into account the tendency to a multi-messenger analysis~\cite{Franckowiak17} of data with its potential for a more accurate exploration of the Universe and mordern trend to open science~\cite{Voruganti17,Wagner15}, it is very important to provide users from geographically distributed locations with access to the data of different astrophysics facilities. 
Today, open access to data or, more generally, open science is becoming increasingly popular.
This is due to the fact that the amount of data received in some experiment often exceeds the capabilities of the relevant collaboration to process and analyze these data.
And only the involvement of all scientists interested in research in this area allows for a comprehensive analysis of the data in full.

Please note that most existing collaborations have a long history and apply methods for data processing they are accustomed to.
So, our approach to the design of  data storage for astroparticle physics should be based on two main principles. 
The first principle is that there is no interference with the existing local storage. 
And the second principle is the processing of user requests in a special service outside the local storage using metadata.
The interaction between local storages and any user of the system should be provided by special adaptors which define a unified interface for data exchange in the system.
Our approach to storage design is based on a single write-multiple read model (SWMR) for accessing raw data or centrally processed data for further analysis. 
The motivation for the solution is that both raw data and data after the initial processing (for example, calibration) should be stored unchanged and presented to users as is upon request.
A similar approach is being discussed in the HDF5 community~\cite{hdf5_swmr}.

The main ideas of the proposed approach are as follows:
\begin{itemize}
\item no changes in the inner structure of local storage;
\item unification of access to local storage based on corresponding adapter modules;
\item use of local data access policies;
\item search of the requested data  using the only metadata on a special service;
\item aggregating the requested data into a new collection and providing the user with access to it;
\item data transfer only at the moment of actual access to them.
\end{itemize}

Based on the above principles and ideas, we propose a concept of distributed storage for astrophysical experiments, which we call APPDS (abbreviated from AstroParticle Physics Distributed Storage). 
The prototype of such distributed storage is developed in the framework of the German-Russian Astroparticle Data Life Cycle Initiative~\cite{Bychkov18}. 
This initiative aims to develop a distributed data storage system in the field of astrophysics of particles by the example of two experiments KASCADE~\cite{KASCADE2} and TAIGA~\cite{TAIGA,TAIGA2}, as well as to demonstrate its viability, stability and efficiency.

Below we discuss the architecture of the distributed data storage and briefly report the current status of the project and the nearest plans.

%%%%%%%%%%%%%%%%%%%%%%%%%%%%%%%%%%%%%%%%%%%%%%%%%%%%%%%%%%%%%%%%%%%%%%%%%%%%%
%%%%%%%%%%%%%%%%%%%%%%%%%%%%%%%%%%%%%%%%%%%%%%%%%%%%%%%%%%%%%%%%%%%%%%%%%%%%%
\section{Architecture of the data storage}
%%%%%%%%%%%%%%%%%%%%%%%%%%%%%%%%%%%%%%%%%%%%%%%%%%%%%%%%%%%%%%%%%%%%%%%%%%%%%
%%%%%%%%%%%%%%%%%%%%%%%%%%%%%%%%%%%%%%%%%%%%%%%%%%%%%%%%%%%%%%%%%%%%%%%%%%%%%

\begin{figure}
%%%%%%%%%%%%%%%%%%%%%%%%%%%%%%%%%%%%%%%%%%%%%%%%%%%%%%%%%%%%%%%%%%%%%%%%%%%%%
\includegraphics[width=\textwidth]{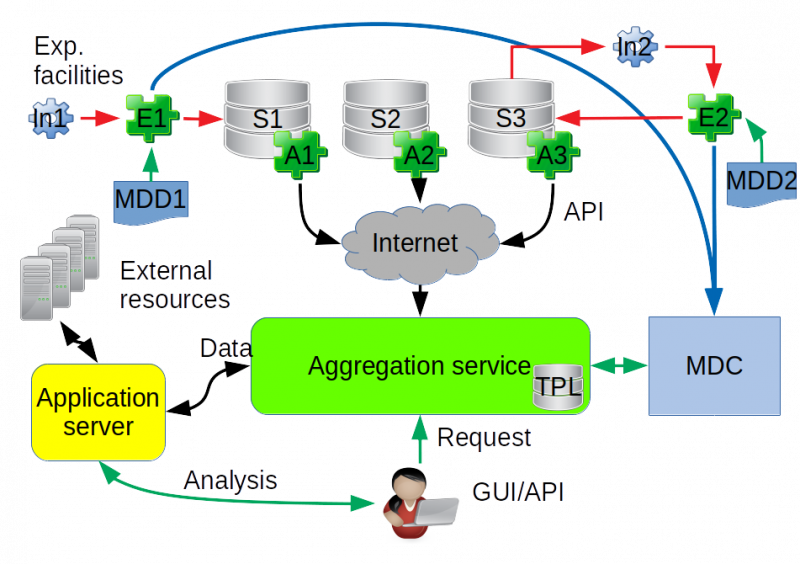}
\caption{The architecture of APPDS} \label{fig1}
%%%%%%%%%%%%%%%%%%%%%%%%%%%%%%%%%%%%%%%%%%%%%%%%%%%%%%%%%%%%%%%%%%%%%%%%%%%%%
\end{figure}

One of the main ideas of the distributed data storage architecture for the physics of astroparticles is that we do not interfere with the work of local storages S1 ... S3 (see Fig.~\ref{fig1}).
This is achieved by using special adapter programs A1 ... A3 that allow local storages to interact with the data aggregation service.
As adapters, we use the CERNVM-FS~\cite{Blomer17} file system to export local file systems to the aggregation service in a read-only mode. 
First, it provides a transparent way for users to interact with local storages. Secondly, the actual transfer of data will only occur when a user actually accesses these data.
Additionally, reducing network traffic can be achieved through the use of CVMFS caching properties.

%%%%%%%%%%%%%%%%%%%%%%%%%%%%%%%%%%%%%%%%%%%%%%%%%%%%%%%%%%%%%%%%%%%%%%%%%%%%%
%%%%%%%%%%%%%%%%%%%%%%%%%%%%%%%%%%%%%%%%%%%%%%%%%%%%%%%%%%%%%%%%%%%%%%%%%%%%%

To retrieve the necessary files, a user forms a request through the web interface provided by the Data Aggregation Service.
When the Data Aggregation Service receives the user request, it requests a response from the Metadata Catalogue~(MDC).
After the Metadata Catalogue responds, the Data Aggregation Service forms the corresponding resulting response and delivers it to the user.

The proposed system offers two types of search conditions for user requests: a file-level search and an event-level search.

In the case of a file-level search, the user requests a set of files, imposing conditions on the metadata (that is, on the data about the
files). 
An example of such a condition is the range of dates of observation of gamma sources in the sky. 
It is important to note that the user will receive in response the corresponding set of files with the same directory structure as in the original repository. 
Thus, the application software can be run without modification, as if the user runs the program locally.

%The second condition is a selection of some events from the files which satisfy the request, for example the energy range of air-shower.
%In this case the events are picked up from the file and aggregated to a new one. 
%The new file is transferred to the user. However, the directory stracture will be preserved too.

In the case of a event-level search, the user wants to select from the files only some events that satisfy the search conditions, for
example, some energy range of the air flow. 
In this case the events are selected from the files and the aggregation service prepares a new one which contains only the necessary events.
The new file is transferred to the user. However, the directory structure will be preserved too.

The processing of user requests is performed by the metadata catalogue, the main purpose of which is to specify, up to an event, in which files and where the data requested by the user are contained. 
The MDC service is built around TimescaleDB \cite{Freedman18,Yang18,Stefancova18}.

The extractors E1, E2 play a key role in the architecture of APPDS.
All data stored in the local storages must pass through the extractors.
The extractors take off metadata from the data and store the metadata in MDC.
The type of the extracted metadata is defined by the metadata description file~(MDD) which is used as input for the extractor. 
The MDD file is written in Kaitai Struct \cite{KaitaiStruct,Bychkov18-2} format with special marks pointing to elements of binary data which are metadata and should be extracted.

The extractor E1 takes out metadata from raw data, while the extractor E2 takes out metadata from centrally processed data (for example, from
data after calibration or calculation of the shower energy).
Thus, the information needed to process user requests is collected in the MDC service.

It is important to note that all services in APPDS are built as microservices \cite{Sill16} and have a well--defined REST API \cite{Fielding00}.  
Some services are running in Docker containers \cite{Docker}.

A more detailed description of the aggregation service and the metadata catalogue service can be found in the papers by M-D.Nguyen~\cite{Nguyen} and I.Bychkov~\cite{Bychkov3} in these proceedings. 

%%%%%%%%%%%%%%%%%%%%%%%%%%%%%%%%%%%%%%%%%%%%%%%%%%%%%%%%%%%%%%%%%%%%%%%%%%%%%
%%%%%%%%%%%%%%%%%%%%%%%%%%%%%%%%%%%%%%%%%%%%%%%%%%%%%%%%%%%%%%%%%%%%%%%%%%%%%
\section{Status}
%%%%%%%%%%%%%%%%%%%%%%%%%%%%%%%%%%%%%%%%%%%%%%%%%%%%%%%%%%%%%%%%%%%%%%%%%%%%%
%%%%%%%%%%%%%%%%%%%%%%%%%%%%%%%%%%%%%%%%%%%%%%%%%%%%%%%%%%%%%%%%%%%%%%%%%%%%%

Currently a prototype of APPDS was deployed in Skobeltsyn Institute of Nuclear Physics, Lomonosov Moscow State University. 
The prototype consists of two local storages interconnected via a local network for modelling distributed storage, an aggregation service and a metadata service based on TimescaleDB.
The next version of the system will also include KCDC~\cite{KCDC} storage at KIT and storage at Irkutsk State University.

Most of the components of the system are written in Python.
As the first-time example of the production use of the system, users of the KASCADE and TAIGA/TUNKA collaborations will gain access to the data of these experiments, as well as the Monte Carlo simulation data. 
It should be mentioned that the system is developed for broad general use and is not limited to astrophysics applications.
\end{document}